\documentclass[journal]{IEEEtran}
\usepackage{graphicx}
\usepackage{amsmath}
\usepackage{textcomp}
\usepackage{array}
\usepackage{multicol}
\usepackage{authblk}
\newcommand*{\rom}[1]{\expandafter\@slowromancap\romannumeral #1@}
\usepackage{longtable}
\usepackage{tabto}
\usepackage{color}
\usepackage[noadjust]{cite}
\usepackage{makecell}
\usepackage{enumitem}
\usepackage{multirow}
\usepackage{bm}

\begin{document}
\title{Remote Sensing Image Scene Classification with Deep Neural Networks in JPEG 2000 Compressed Domain%
\thanks{This work is funded by the European Research Council (ERC) through the ERC-2017-STG BigEarth Project under Grant 759764.}%
}

\author{Akshara~Preethy~Byju,
        Gencer~Sumbul,~\IEEEmembership{Graduate Student Member,~IEEE},
        Beg\"um~Demir,~\IEEEmembership{Senior Member,~IEEE},
        and~Lorenzo~Bruzzone,~\IEEEmembership{Fellow,~IEEE}
\thanks{Akshara Preethy Byju and Lorenzo Bruzzone are with the Department of Information Engineering and Computer Science, University of Trento, Trento, Italy (e-mail: \mbox{akshara.preethybyju@unitn.it}; \mbox{lorenzo.bruzzone@unitn.it}).}
\thanks{Gencer Sumbul and Begüm Demir are with the Faculty of Electrical Engineering and Computer Science, Technische Universit\"at Berlin, Berlin, Germany (e-mail: \mbox{gencer.suembuel@tu-berlin.de}; \mbox{demir@tu-berlin.de}).}
}


\maketitle

\begin{abstract}
To reduce the storage requirements, remote sensing (RS) images are usually stored in compressed format. Existing scene classification approaches using deep neural networks (DNNs) require to fully decompress the images, which is a computationally demanding task in operational applications. To address this issue, in this paper we propose a novel approach to achieve scene classification in JPEG 2000 compressed RS images. The proposed approach consists of two main steps: i) approximation of the finer resolution sub-bands of reversible biorthogonal wavelet filters used in JPEG 2000; and ii) characterization of the high-level semantic content of approximated wavelet sub-bands and scene classification based on the learnt descriptors. This is achieved by taking codestreams associated with the coarsest resolution wavelet sub-band as input to approximate finer resolution sub-bands using a number of transposed convolutional layers. Then, a series of convolutional layers models the high-level semantic content of the approximated wavelet sub-band. Thus, the proposed approach models the multiresolution paradigm given in the JPEG 2000 compression algorithm in an end-to-end trainable unified neural network. In the classification stage, the proposed approach takes only the coarsest resolution wavelet sub-bands as input, thereby reducing the time required to apply decoding. Experimental results performed on two benchmark aerial image archives demonstrate that the proposed approach significantly reduces the computational time with similar classification accuracies when compared to traditional RS scene classification approaches (which requires full image decompression).
\end{abstract}
.

\section{Introduction}
\IEEEPARstart{A}{dvancements} in satellite technologies have led to a huge increase in the volume of remote sensing (RS) image archives up to an unprecedented level. Subsequently, the extent of information that can be inferred from these massive archives is also increasing, enabling applications. Considering the volume of such massive archives as well as the complexity of RS images, developing efficient scene classification methods is one of the most important research topics in RS. 

Scene classification aims to assign a class label to a given RS image based on the analysis of descriptors that characterize its semantic content. The performance of any scene classification method mainly depends on its ability to characterize a given RS image with efficient feature representations. In the last decade, several handcrafted features were introduced in the RS literature \cite{Zhang2017, Georgescu2016, Puetz2006,Dong2019 ,Tekeste2018,Hu2015,Zhang2015,Zhu2016}. However, exploiting the aforementioned handcrafted features for scene classification tasks in massive RS image archives that contain petabytes of data is inefficient, time-demanding and computationally-complex. Moreover, the discriminative power of these handcrafted features is often shallow and requires human intelligence, which incurs additional labor costs. To address these limitations, in recent years several deep learning (DL) algorithms were introduced in RS community~\cite{Chen2016, Chen2014, Padmanabhan2016, Scott2017, Sevo2016}. Among them convolutional neural networks (CNNs) have demonstrated their remarkable ability to learn the high-level semantic content of RS image resulting in high classification accuracies. These networks hierarchically learn the intrinsic patterns within the images through several convolution and pooling operations to obtain distinctive feature descriptors. Several studies have shown very good performance of CNNs in the RS domain for several applications, including scene classification~\cite{Dong2019}, \cite{Chen2016}, \cite{Cheng2018}. In the early years, CNN models were trained from scratch with a considerable amount of training data. However, recent experiments have shown that using pretrained DL networks (e.g. ImageNet~\cite{Takamitsu1978}, VGG16~\cite{Zhang2016a}, GoogLeNet~\cite{Zeng2016}, CaffeNet~\cite{Jia2013}) in RS  domain has remarkably improved classification performance. In~\cite{Xia2017a}, it is shown that the classification accuracy of a CNN model with a considered pre-trained network outperforms what obtained when a model is trained from scratch. This mainly depends on the small number of training samples usually available in RS applications.

Recently, several studies were conducted in order to enrich the discriminative power of the image descriptors obtained through conventional CNN models. As an example, combining features obtained at multiple resolutions to represent a scene provides significant improvement in classification accuracies~\cite{Duarte2018}. Although this approach is rotation as well as translation invariant, it is time-demanding to train images at several resolutions using a simple CNN model. Zheng et al.~\cite{Zheng2019} proposed a deep representation where the  multiple-scale features are obtained from the image feature maps using multiscale pooling (MSP) to improve the classification performance at a faster rate. To improve the classification performance, features obtained from relevant image areas can be considered. This can be done by introducing the attention mechanisms. In~\cite{Sumb2019}, local image descriptors are obtained with  recurrent attention to perform multi-label RS scene classification. Attention mechanism avoids irrelevant areas and focus to obtain features from relevant image regions thereby reducing the number of parameters as well as the computational time required to perform a specific task. Guo et. al.~\cite{Guo2019} have proposed a network where both local and global features are learned through a local global attention framework. 

To mitigate the problem of overfitting that generally arises due to the limited availability of annotated data, generative adversarial networks (GANs) became popular to address scene classification problems in RS. GANs learn the hidden structure in the given input data and consists of a generator (that learns the semantic contents of the data) and a discriminator (that classifies the generated and input images)~\cite{Yu2019, Radford2016, Roy2018b, Lin2017, Duan2019, Zhu2018}. MARTA-GAN proposed in~\cite{Lin2017}, was one among the first efforts to learn feature representations to perform unsupervised aerial scene classification. The performance of GAN models depends on the quality of the modelling of the structure of generated images. In~\cite{Duan2019}, GAN-NL has proposed to effectively model the non-local dependencies in the generated images and has shown remarkable improvements in the classification accuracies. Roy et al.~\cite{Roy2018b} proposed a semantic fusion GAN, where the feature representations are obtained using a standard Deep Convolutional GAN (DCGAN) combined with an external deep network. In~\cite{Zhu2018}, two CNNs are integrated that serve as the generator and discriminator model of GAN to classify hyperspectral images. The authors proposed 1D-GAN and 3D-GAN that are used to classify the spectral and spatial-spectral information, respectively. Although the current state-of-the-art GAN networks demonstrate an improvement in classification performance, their optimization and training are time demanding.

To reduce the storage required for huge amounts of data, RS images are compressed before storing them in the archives~\cite{Zhou2015}. Several image compression algorithms such as Differential Pulse Code Modulation (DPCM), Adaptive DPCM (ADPCM), Joint Photographic Experts Group (JPEG), JPEG 2000 were introduced in RS~\cite{Zhou2015}. Among several compression algorithms, JPEG 2000~\cite{Taubman} became very popular in RS due to its multiresolution paradigm, scalability and high compression ratio. JPEG 2000 algorithm is used to compress RS images acquired by most of the recent applied satellites (such as Sentinel-2 and PRISMA) in their archives. Thus, before performing any scene classification of images in compressed archives, the image decompression task should be performed. This is computationally-demanding and ineffective when considering real large-scale RS image archives that may contain petabyte scale data. Considering the complexity and size of RS image, the amount of time required to decompress any image cannot be neglected in real large-scale RS image archives. 

To address the aforementioned limitation, in this paper, we propose a novel approach based on DNNs for scene classification of compressed RS images. We assume that the images in the archive are compressed using JPEG 2000. The proposed approach aims to minimize the amount of decompression required for the classification of compressed RS images while maintaining similar accuracies when compared to the conventional state-of-the-art scene classification approaches. To achieve this, the proposed approach consists of two main steps: i) approximation of the finer (higher) resolution wavelet sub-bands used in JPEG 2000 compression algorithm; and ii) feature extraction and scene classification of the approximated finer resolution wavelet sub-bands. The proposed approach initially obtains the codestreams associated with the coarsest resolution wavelet sub-bands of the considered JPEG 2000 compressed RS image. Then, in the first step, finer resolution wavelet sub-bands (image) are approximated through a series of transposed convolutional layers. The second step aims to obtain the features associated with the approximated finer resolution wavelet sub-bands and perform scene classification based on the learnt descriptors. This is obtained by introducing a loss function that learns the parameters associated with both approximation and scene classification in an end-to-end unified neural network. In the classification phase, the proposed approach requires only the codestreams associated with the coarsest wavelet resolution sub-bands and thereby reduces the time required to preform decompression of the images. The proposed approach explores the hierarchical multiresolution feature space in a unified framework and achieves optimal resource utilization for scene classification. The effectiveness of the proposed approach was evaluated by using two different aerial benchmark image archives: NWPU-RESISC45~\cite{Cheng2017a} and AID~\cite{Xia2017a}. Please note that the aim of this study is not to introduce a compression algorithm but to propose a novel DL approach that requires minimally decoded wavelet subband information to obtain optimal classification, thereby significantly minimizing the computational cost of the decoding step.

The remainder of this paper is organized into five sections: Section II gives a brief review of the related works. Section III explains the proposed approach. Section IV describes the data sets and the experimental setup, while Section V illustrates the experimental results with discussion. Finally, Section VI draws the conclusion of the work.
\begin{figure*}[!t]
\centering
	\includegraphics[width=\textwidth]{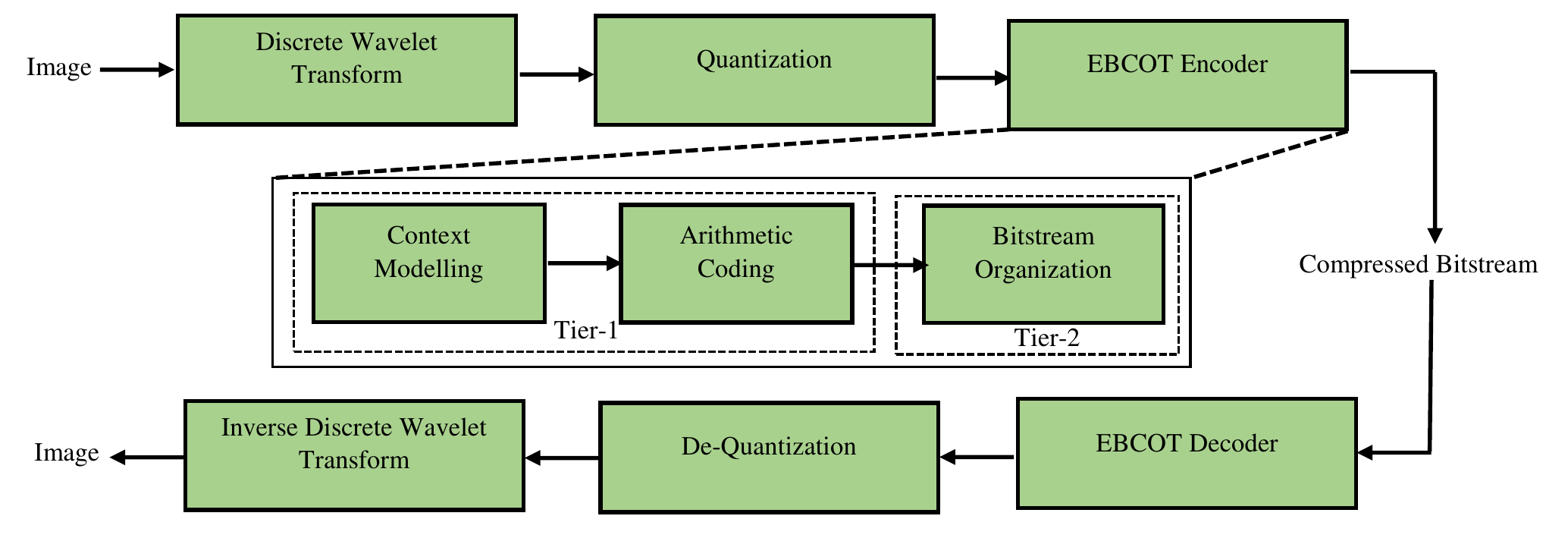}
	\caption{General block scheme of the JPEG 2000 compression and decompression algorithm.}
\end{figure*}

\section{RELATED WORKS}

\subsection{JPEG 2000 Algorithm}
JPEG 2000 is one among the most popular compression algorithms and image coding systems that rely on the wavelet-based approach. Figure 1 illustrates the block scheme in the encoding and decoding of JPEG 2000 compression algorithm. It includes: i) Discrete Wavelet Transform (DWT); ii) Quantization; and iii) Embedded Block Coding with Optimized Truncation. DWT when applied to a given image helps to achieve multiresolution image description and decomposes the given image into one approximation (LL) sub-band and three detail sub-bands (LH, HL and HH). The LL sub-band is further decomposed to approximation and detail sub-bands, if one requires more than one decomposition level. The wavelet coefficients of each sub-band are quantized using a step-size quantizer as selected by the user. As we consider lossless compression for the images in the archive, we do not perform quantization and thus this step is neglected. The sub-bands are then sub-divided into rectangular blocks referred as \textit{precinct} and each \textit{precinct} is again subdivided into \textit{codeblocks} that are usually of size $64\times64$. The minimum size of a \textit{codeblock} should be $32\times32$. These \textit{codeblocks}, which are represented using \textit{bitplanes}, are then entropy coded using the EBCOT algorithm. Entropy coding in JPEG 2000 compression algorithm is separated into \textit{Tier-1} and \textit{Tier-2} coding. In \textit{Tier-1} coding, each \textit{codeblock} is coded using i) \textit{Context Modelling} and ii) \textit{Arithmetic Coding}. The probability of occurrence of a particular bit can be determined from the contextual information, which is obtained from three passes: \textit{significance propagation pass}, \textit{refinement pass} and \textit{cleanup pass}. In \textit{Tier-2} coding, the bitstreams obtained after \textit{arithmetic coding} are organized into several \textit{packets} and \textit{layers}. A given \textit{packet} includes the codestream associated with a particular \textit{precinct} and allows to access data based on the \textit{quality}, \textit{resolution} or \textit{band}. This packet structure organization allows to access progressively and hierarchically to the information of a given image.

\subsection{JPEG 2000 based Feature Descriptors}
Features that can be obtained from JPEG 2000 compressed images for scene classification are broadly divided into two categories: header-based and wavelet-based features. Header-based features are directly obtained from the bitstreams of a JPEG 2000 compressed image, where one can acquire information such as the number of bytes $B$ used to entropically encode a given image or the maximum number of significant bitplanes $MB$ in a given codeblock~\cite{Descampe2011}. In~\cite{Mallat1992b}, Mallat illustrates that these features represent the singularities present in an image, which can be utilized for image classification/retrieval.

Wavelet-based features are obtained after the partial decompression of a given compressed image. They efficiently model spectral, texture and shape information obtained from the approximation and detail (horizontal, vertical, diagonal) wavelet sub-bands. The problem of obtaining features from wavelets has been studied extensively during the last decade. By observing that the detail sub-bands have a near-Gaussian behavior, most of the early research works on image classification/retrieval focused on modelling the detail wavelet sub-bands using Generalized Gaussian Distribution (GGD)~\cite{Allili2012}, Gaussian Mixture Model (GMM)~\cite{Wang2014}, Generalized Gamma Density (GTD)~\cite{Choy2010} and their variants. Although these statistical representations are highly discriminative, they are computationally-complex and time-consuming.  The energy and mean descriptor obtained from the detail sub-bands were also used as the texture features~\cite{Pi2006}. In addition, features extracted by the co-occurrence matrix (i.e. contrast, homogeneity, energy, variance, correlation) were considered to model the texture features. The histogram obtained from the probability of joint distribution of bitplanes that are obtained directly from the JPEG 2000 codestreams have been used for image retrieval~\cite{Pi2006}. In~\cite{teynor}, the spectral information obtained from the approximation sub-band exploited to perform image retrieval in a JPEG 2000 compressed archive. In~\cite{DeVes2014}, the moduli as well as angle of the wavelet coefficients obtained from the horizontal and vertical wavelet sub-bands were found very effective in modelling the edge features for image classification. In~\cite{Kekre2010}, morphological operations such as dilation and erosion applied to wavelet sub-bands were used to obtain shape based features to perform image retrieval from JPEG 2000 compressed images. However, these handcrafted features are unable to capture the high-level semantic information when compared to deep features. Thus this work focuses on developing a novel approach that benefits from DNNs to achieve efficient scene classification performance in JPEG 2000 compressed image archives.

\begin{figure*}[!t]
\centering
	\includegraphics[width=\textwidth]{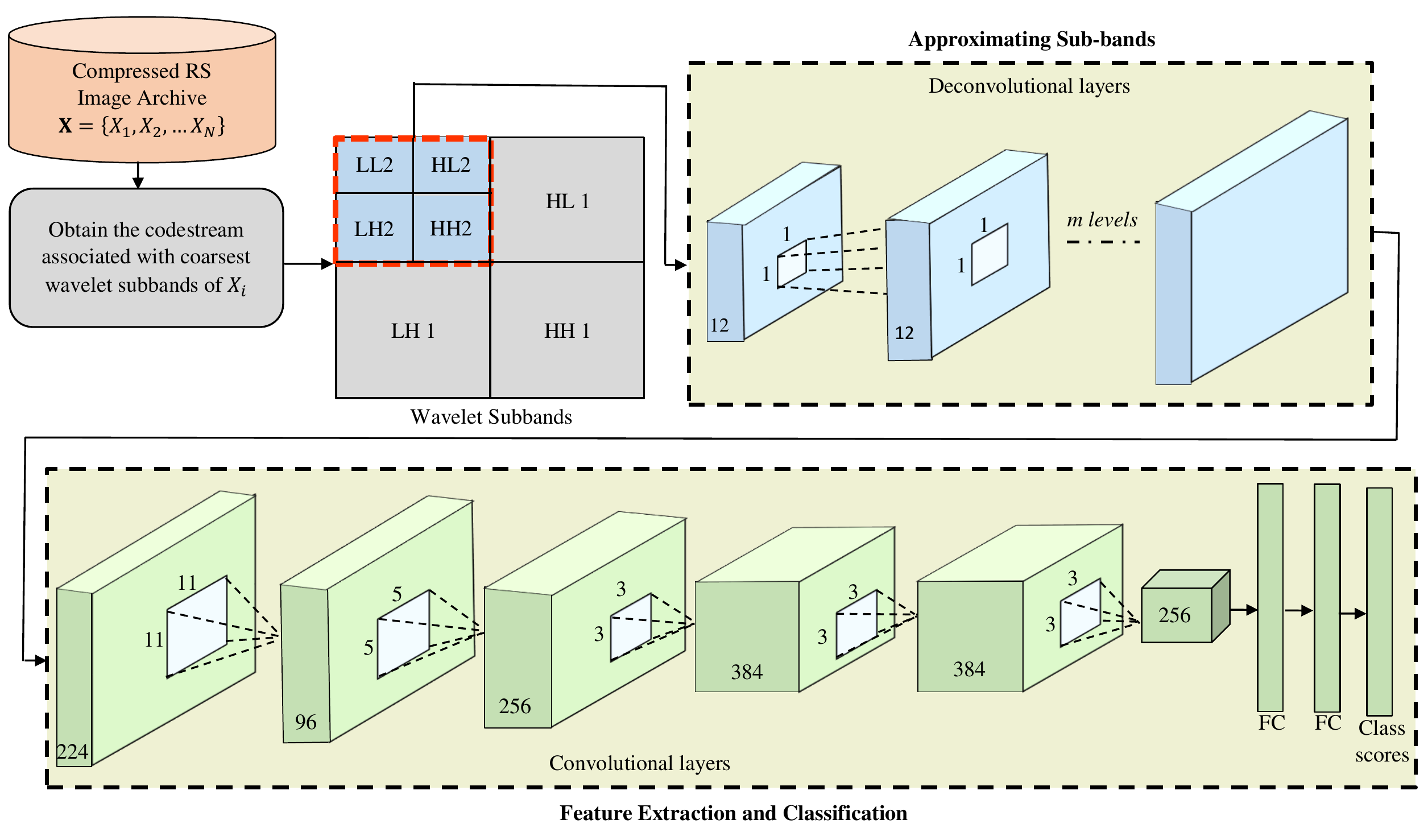}
	\caption{Block scheme of the proposed approximation approach in the compressed domain.}
\end{figure*}

\section{PROPOSED SCENE CLASSIFICATION APPROACH IN THE JPEG 2000 COMPRESSED DOMAIN}
\subsection{Problem Formulation}

Let $\textbf{X}=\{X_i\}^\textit{N}_{i=1}$ be an archive that contains \textit{N} JPEG 2000 compressed RS images, where $X_i$ represents the $i^{th}$ image. The main objective of the proposed approach is to assign a class label $y_i\in \textbf{Y}$, where $\textbf{Y}$ is a set of class labels, to a given input image $X_i\in\textbf{X}$. Let us assume that all the images in the archive are decomposed up to $L$ resolutions. Each image in the archive will be represented as one approximation sub-band and $3L$ detail sub-bands (i.e. horizontal, vertical and diagonal). In JPEG 2000 compressed image archive, the straightforward approach to perform scene classification is to: i) apply entropy decoding to the code streams associated with all the images in the archive; and ii) obtain the image descriptors. However, decoding all the images from a compressed archive is time-demanding and computationally-expensive. Thus, we propose a novel approach based on DNNs that efficiently approximates a decompressed image to perform scene classification in a large scale JPEG 2000 compressed image archive. Our objective is: i) to implement a novel DL approach that performs scene classification in the compressed domain with minimal decompression; and ii) to reduce the computational time when compared to models that require fully decoded images. To achieve this, the proposed approach consists of two main steps: i) approximation of the finer (higher) resolution wavelet sub-bands used in JPEG 2000 compression algorithm; and ii) feature extraction and scene classification of the approximated finer resolution wavelet sub-bands. Figure 2 shows the block scheme of the proposed approach and each step is explained in the following subsections. 

\subsection{Approximation of the Wavelet Coefficients}
This step aims to approximate the finer (higher) resolution wavelet sub-bands (or the image itself) through a series of transposed convolutional layers. To this end, the proposed approach considers $m$ transposed convolutional layers, where $m$ corresponds to the number of wavelet decomposition levels that were initially used to compress a given image $X_i\in \textbf{X}$. 
Given a compressed image $X_i$, we initially obtain the codestream associated with the coarsest level wavelet sub-band (see Fig. 2) that provides the global scale information of any given image. 
Let $G^L=\{a^L_{X_i}$, $h^L_{X_i}$, $v^L_{X_i}$, $d^L_{X_i}\}$ denote the approximation, horizontal, vertical and diagonal sub-bands of an image $X_i$ at the $L^{th}$ wavelet decomposition level (coarsest wavelet sub-band). Let $A^{L-1}=\{a^{L-1}_{X_i}$, $h^{L-1}_{X_i}$, $v^{L-1}_{X_i}$, $d^{L-1}_{X_i}\}$ be the finer level approximated sub-bands at level $L-1$. 
\begin{figure}
\centering
	\includegraphics[width=0.5\textwidth]{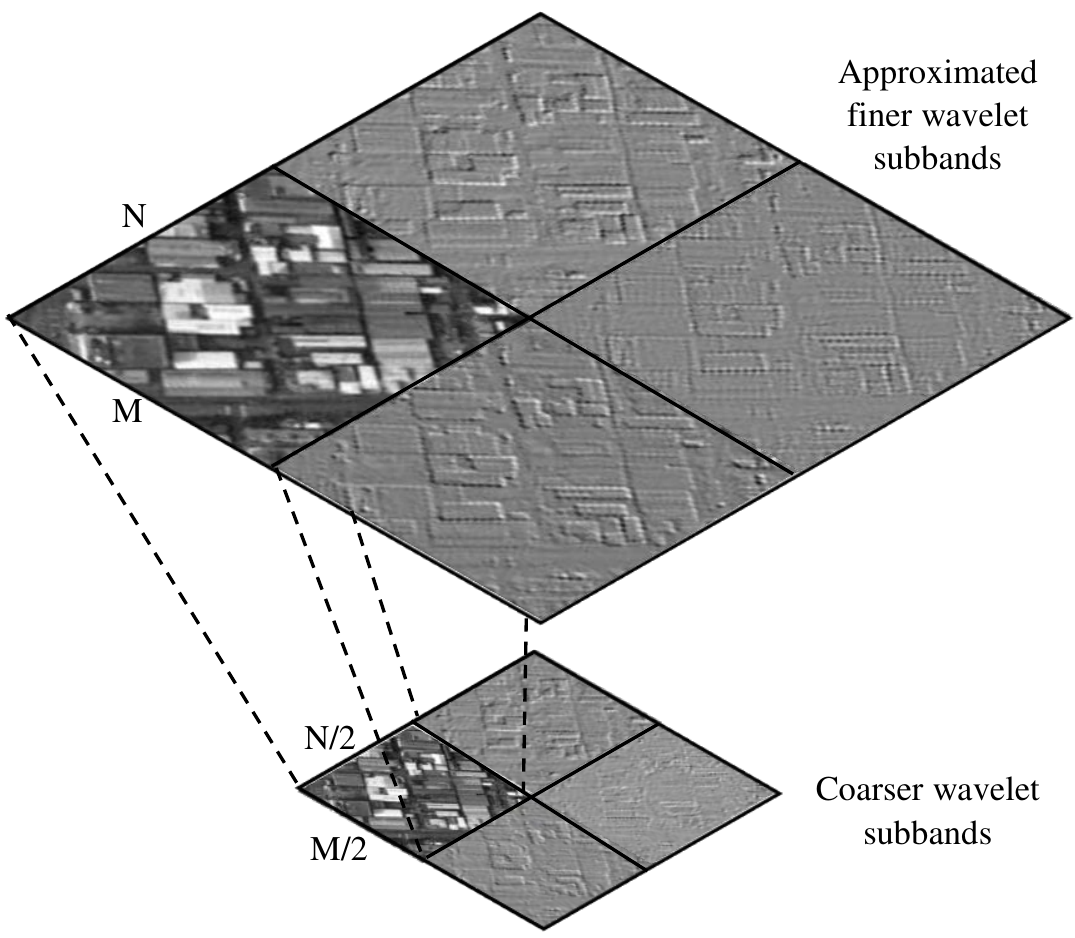}
	\caption{Example of approximation of a finer level wavelet sub-bands using transposed convolution.}
\end{figure}
\begin{figure*}[!t]
\centering
	\includegraphics[width=\textwidth]{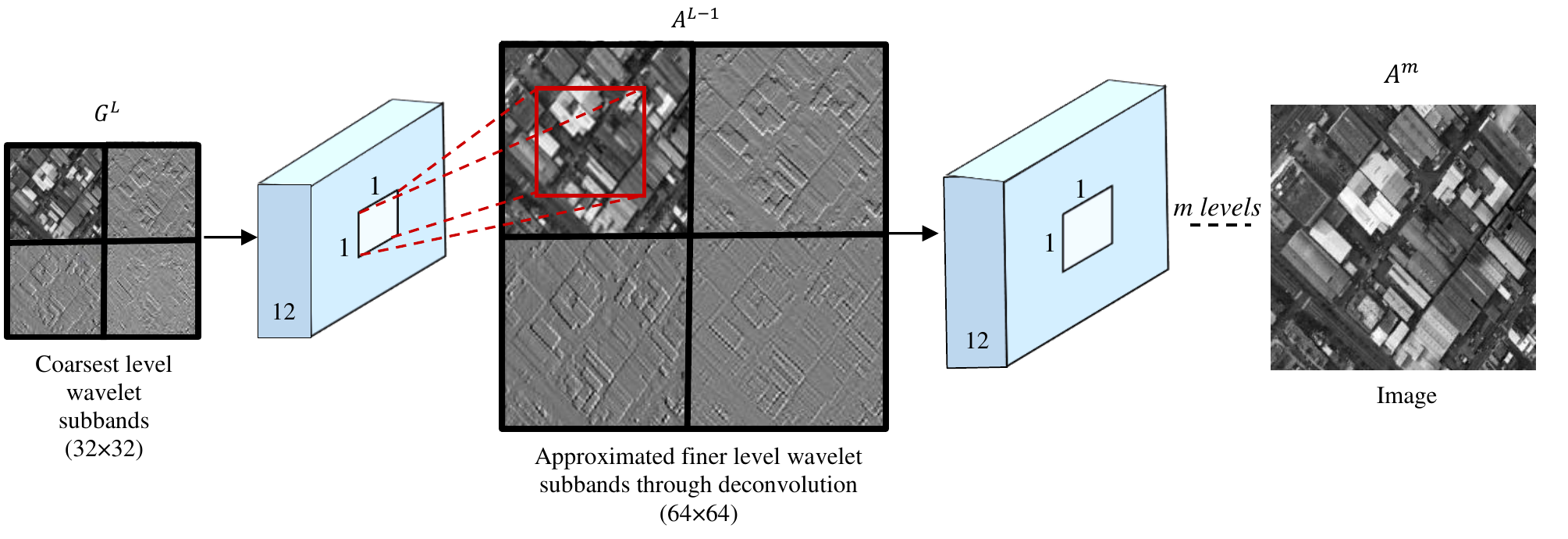}
	\caption{Illustration of the proposed approach when the approximated image is obtained from codestream of coarsest level wavelet sub-band. }
\end{figure*}
\begin{figure*}
\centering
	\includegraphics[width=\textwidth]{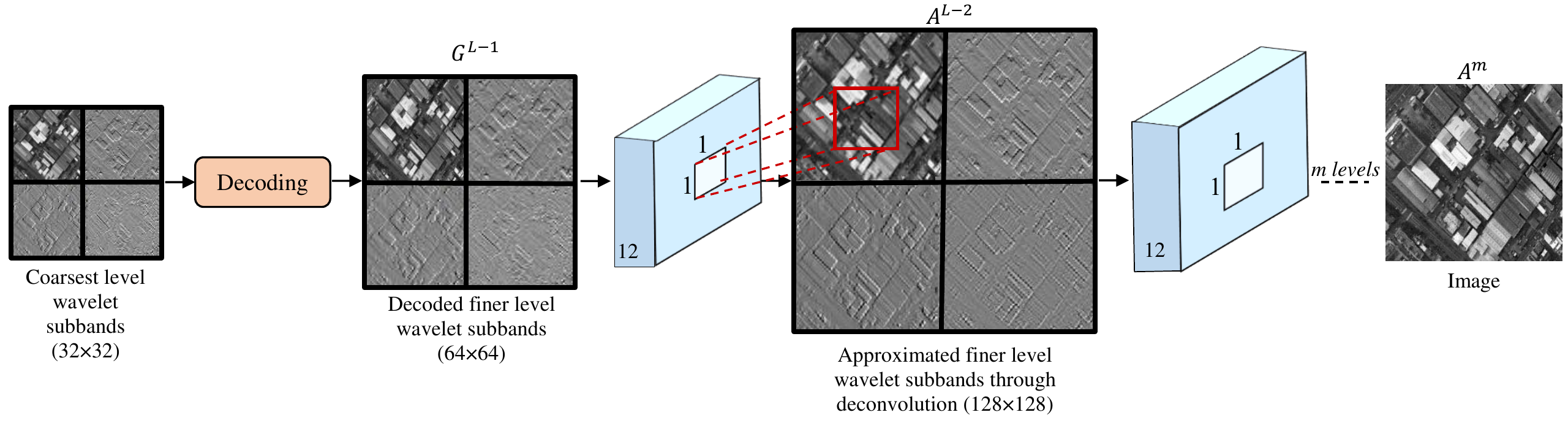}
	\caption{Illustration of the proposed approach when the approximated image is obtained from codestream of finer level wavelet sub-band.}
\end{figure*}

We consider \textit{'convolution'} operation as a matrix multiplication between the flattened input and a sparse matrix $C$. As an example, $G^L$ can be obtained by applying convolution operation on $A^{L-1}$ as follows:
\begin{equation}
	G^L={C\cdot A^{L-1}}.
\end{equation}
The non-zero elements in the sparse matrix $C$ can be constructed using the kernel coefficients of the convolution operation as follows:
\begin{equation} 
C=
\left[\begin{smallmatrix}
           &        &        &        &         &        &        &        &        &    \\
    k_{11} &  ...   & k_{1q} & 0      &  ...    & k_{2q} & ...    & k_{pq} & 0      & ...\\
           &        &        &        &         &        &        &        &        &    \\
    0      & k_{11} &  ...   & k_{1q} & 0       & ...    & k_{2q} & ...    & k_{pq} & ...\\
           &        &        &        &         &        &        &        &        &    \\
    0      & 0      & k_{11} &  ...   & k_{1q}  & 0      & ...    & k_{2q} &        & ...\\
           &        &        &        &         &        &        &        &        &    \\
           &        &        &        &         &        &        &        &        &    \\
    ...    &        &        &        &         &        &        &        &        & ...\\
           &        &        &        &         &        &        &        &        &    \\
    0      & 0      & 0      &  0     &         &        &  ...   &        &        & k_{pq}\\
           &        &        &        &         &        &        &        &        &    \\
\end{smallmatrix}\right]
\end{equation}
 where $p$ and $q$ represent the kernel size and $k_{ij}$ is the element of the kernel (where $i$ and $j$ are the row and column indices of the kernel, respectively). Convolution operation takes the input matrix $A^{L-1}$ which is then flattened into a vector and multiplies the flattened input with $C$. The matrix multiplication result is reshaped to obtain the final output $G^L$. It is worth noting that during the forward and backward passes of CNNs, convolution operations are applied with $C$ and $C^T$, respectively. In computer vision, \textit{transposed convolution} has proved to be an efficient algorithm that uses the gradient of the convolution operation (for a given image) to perform image restoration and reconstruction~\cite{Zhang2018}. Our proposed approach approximates the finer wavelet sub-bands using transposed convolution as shown in Fig. 3. Given a kernel, the transposed convolution operation multiplies the flattened input vector $G^L$ with $C^T$ during the forward pass and multiplies with ${(C^T)^T = C}$ during the backward pass to obtain $A^{L-1}$. The finer level wavelet sub-bands can be obtained as:
\begin{equation}
	A^{L-1}={C^{T} \cdot G^L},
\end{equation}
where we swap the backward and forward passes of the convolution operation which is used in standard CNNs. Accordingly, using $m$ transposed convolutional layers, the proposed approach allows to approximate the image $A^m$. For the transposed convolutional layers, if we use a stride $S$, padding $P$ and kernel size $k$, then the size of the approximated wavelet sub-bands ($A^{L-1}_{size}$) obtained from the coarser level wavelet sub-bands ($A^{L}_{size}$) is given by:
\begin{equation}
    A^{L-1}_{size} = S * (A^L_{size}-1) + k - 2P.
\end{equation}
The proposed approximation approach reflects the inherent multiresolution paradigm within the JPEG 2000 compression algorithm within an end-to-end unified framework. While approximating sub-bands, we consider two scenarios.
\subsubsection{Scenario 1: Minimal Decoding}
In this scenario, the proposed approach obtains only the codestreams associated with the coarsest level ($L^{th}$ level) wavelet sub-bands to approximate the finer level sub-bands (image itself). Here, the aim is to minimize the amount of decompression time required to perform scene classification by approximating wavelet sub-bands (image) using only the coarsest level sub-band. The coarsest level wavelet sub-band provides global scale information of the considered image. Thus, in this scenario, although the amount of time required for decompression is significantly reduced, the quality of approximation is moderately diminished. Fig. 4 illustrates the case when the proposed approach takes the codestreams associated with $32\times32$ coarsest level wavelet sub-band to approximate the image $A^m$ using $m$ transposed convolutional layers.

\subsubsection{Scenario 2: Partial Decoding}
In this scenario, the proposed approach takes the coarsest level ($L^{th}$ level) wavelet sub-band information to decode the finer level ($L-1^{th}$ level) wavelet sub-band, which is exploited to approximate the finest level wavelet sub-bands (image itself). In this scenario, the amount of required decompression time is reduced moderately to achieve favourable performance, when compared to the case where the images require full decompression. The finer level wavelet sub-bands provide fine scale information of a given image. Thus, the wavelet sub-bands (image) approximated from the finer level sub-bands incorporate the detailed fine scale information that enhance the classification accuracy with moderate reduction in time. Fig. 5 illustrates the case when a $32\times32$ coarsest level wavelet sub-band is employed to decode the finer level sub-bands of size $64\times64$. Then, deconvolution is applied to the decoded finer level wavelet sub-bands to approximate the finest level wavelet sub-bands (image). 

\subsection{Feature Extraction and Classification}
The feature extraction and classification step aims to obtain features from the approximated wavelet sub-bands (image). To this end, we consider a neural network with five convolutional layers with a number of filters similar to that of the AlexNet~\cite{Gonzalez2007} and two fully connected (FC) layers. By modifying the feature extraction and classification steps, we can obtain powerful discriminative features. To demonstrate the effectiveness of recent DL models when used in the compressed domain wavelet subband information, we selected the ResNet50~\cite{He2016} architecture to compare with the results obtained by the AlexNet. Then, the output obtained from the final FC layer is mapped into classification scores. To reduce information loss, we considered zero padding and stride of 1 in each convolutional layer, which is followed by a max-pooling layer except the third and fourth. 

The total loss ($\mathcal{L}_{total}$) of the proposed approach is the sum of the approximation loss ($\mathcal{L}_{approximation}$) and classification loss ($\mathcal{L}_{classification}$), which is obtained as:
\begin{equation}
    \mathcal{L}_{total} =  \mathcal{L}_{classification} + \mathcal{L}_{approximation}.
\end{equation}
The approximation loss $\mathcal{L}_{approximation}$ is obtained by calculating the sum of mean squared errors (MSE) between the approximated wavelet sub-bands and decoded wavelet sub-bands at each level $l$ as follows:
\begin{equation}
	\mathcal{L}_{approximation}=\sum_{i=L}^{1}\sum_{j=1}^{M}\sum_{k=1}^{N}{||A^i(w[j,k])-D^i(w[j,k])||}^2
\end{equation}
where $M \times N$ represents the size of the considered wavelet sub-bands at level $l$, $w[j,k]$ denote the wavelet coefficient at position [$j,k$] and $D^i$ represents the decoded wavelet sub-band at any given level $i$. 
The classification loss $\mathcal{L}_{classification}$ is the cross-entropy loss function, which is predominantly used for scene classification problems and defined as follows:
\begin{equation}
    \mathcal{L}_{classification} =-\sum_{i=1}^{Q}y_{i}log \hat{y}_{i}
\end{equation}
where $\hat{y}_{i}$ denote the predicted class label. To improve the classification performance, batch normalization (BN) and dropout were carried out after each convolutional layer. To overcome vanishing gradient problem, Rectified Linear Unit (ReLU) activation was used after both the convolutional and transposed convolutional layers. Section IV provides the more detailed information regarding the training details and the parameters. 

\section{DATASET DESCRIPTION AND EXPERIMENTAL SETUP}
Several experiments were performed to evaluate the performance of the proposed approach on two benchmark archives. The first one is the NWPU-RESISC45~\cite{Cheng2017a} benchmark archive that consists of 31,500 images associated with 45 different categories (i.e. airplane, airport, baseball diamond, basketball court, beach, bridge, chaparral, church, circular farmland, cloud, commercial area, dense residential, desert, forest, freeway, golf course, ground track field, harbor, industrial area, intersection, island, lake, meadow, medium residential, mobile home park, mountain, overpass, palace, parking lot, railway, railway station, rectangular farmland, river, roundabout, runway, sea ice, ship, snowberg, sparse residential, stadium, storage tank, tennis court, terrace, thermal power station and wetland). Each category has 700 scene classes and each image in the archive has the size of $256\times256$ with a varying spatial resolution between 0.2m to 30m per pixel. The reader is referred to~\cite{Cheng2017a} for detailed information.

The second archive is the AID~\cite{Xia2017a} benchmark archive that contains 10,000 images associated with 30 different categories (i.e. airport, bare land, baseball field, beach, bridge, center, church, commercial, dense residential, desert, farmland, forest, industrial, meadow, medium residential, mountain, park, parking, playground, pond, port, railway station, resort, river, school, sparse residential, square, stadium, storage tanks, viaduct). Each image has the size of $600\times600$ pixels with a spatial resolution in the range from 0.5m to 8m. For more detailed information, the reader is referred to~\cite{Xia2017a}. 

\begin{table}
	\caption{NUMBER OF IMAGES CONSIDERED FOR EACH ARCHIVE IN THE TRAINING, VALIDATION AND TEST DATA.}
    \centering
	\begin{tabular}{ |c| c| c| c|}
    \hline
    \centering
     Image Archive  &   Training   &    Validation & Test   \\
     \hline
  NWPU-RESISC45 & 25200 & 3150&3150 \\
     \hline
  AID & 8000 & 1000&1000 \\
  \hline
\end{tabular}
\end{table}
To assess the effectiveness of the proposed model, the images of both archives were compressed using the JPEG 2000 algorithm. TensorFlow deep learning library was utilized for the proposed model. Due to the minimum codeblock size constraint (see Section II-A), we considered a three level wavelet decomposition for both image archives ($L=3$). The codestreams associated with the coarsest wavelet sub-band ($l=3$) is used as the input to the proposed approach. The number of transposed convolutional layers ($m$) is equivalent to the number of wavelet decomposition levels used in the considered image archive. To avoid information loss, we selected the size of the filter as $1\times1$ with stride $1$ and padding $0$. The number of filters used for approximating the wavelet sub-bands is $12\times12$ and the image is $3\times3$. For scenario 2, we considered decoding up to $(m-1)$ wavelet decomposition levels. Both image archives were initially divided into three subsets: training (80\%), validation (10\%) and test (10\%) as shown in Table I. Images included in each subset were randomly sampled. The training of the proposed approach was carried out with the Stochastic Gradient Descent (SGD), which uses the Adaptive Moment Estimation (Adam). During training, the Xavier initialization method was used for the parameter initialization. As there is no pre-trained models to perform scene classification in the compressed domain, all the experiments were performed starting from scratch. In addition, to achieve accurate performance, experiments were carried out varying learning rate between 0.1 to 0.0001. The performance of the proposed architecture was assessed quantitatively and qualitatively by using: 1) classification accuracy; 2) computational time (in sec) of training, validation and test phases; and 3) Root Mean Square Error (RMSE) of the approximated sub-band images. It is worth noting that computational time of the test phase was considered as classification time. All the experiments were performed with Nvidia Tesla V100. 
\begin{figure*}
\centering
		\includegraphics[scale= 0.50]{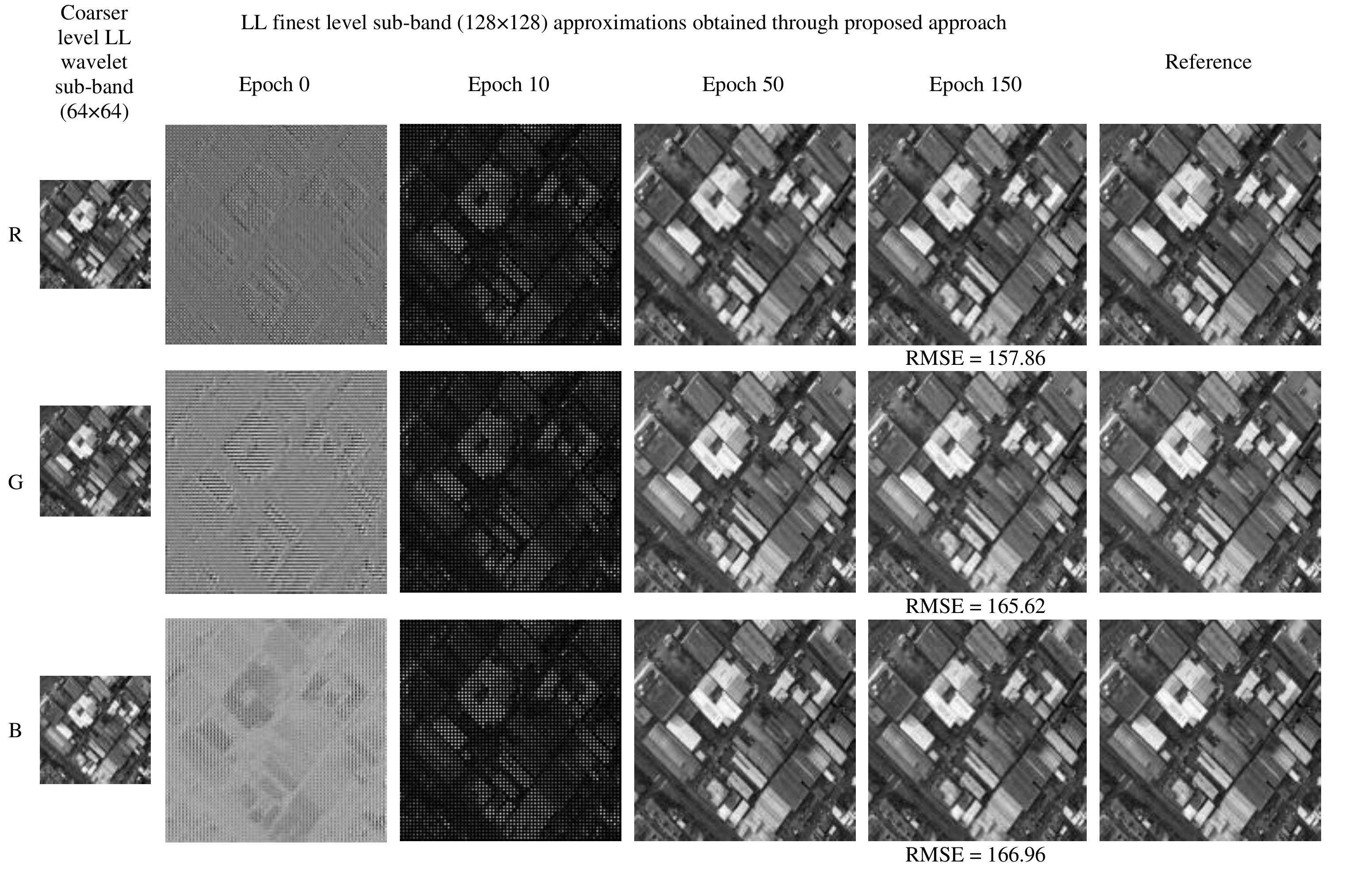}
	\caption{Qualitative results of sub-band approximations associated to LL wavelet sub-band of an image belonging to building category (NWPU-RESISC45 archive).}
\end{figure*}
\begin{figure*}
\centering
		\includegraphics[scale= 0.50]{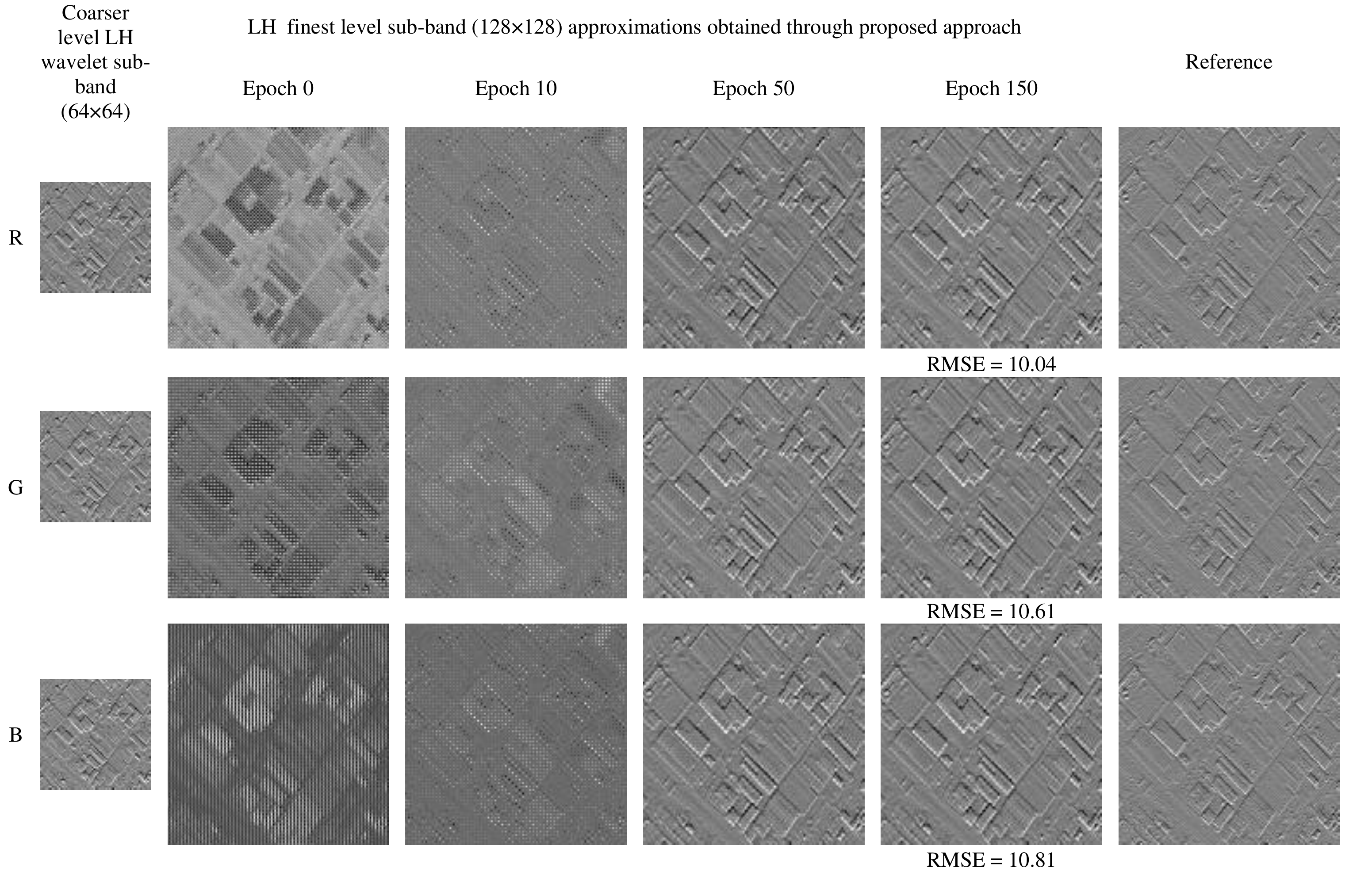}
	\caption{Qualitative results of sub-band approximations associated to LH wavelet sub-band of an image belonging to building category (NWPU-RESISC45 archive).}
\end{figure*}
\begin{figure*}
\centering
		\includegraphics[scale= 0.50]{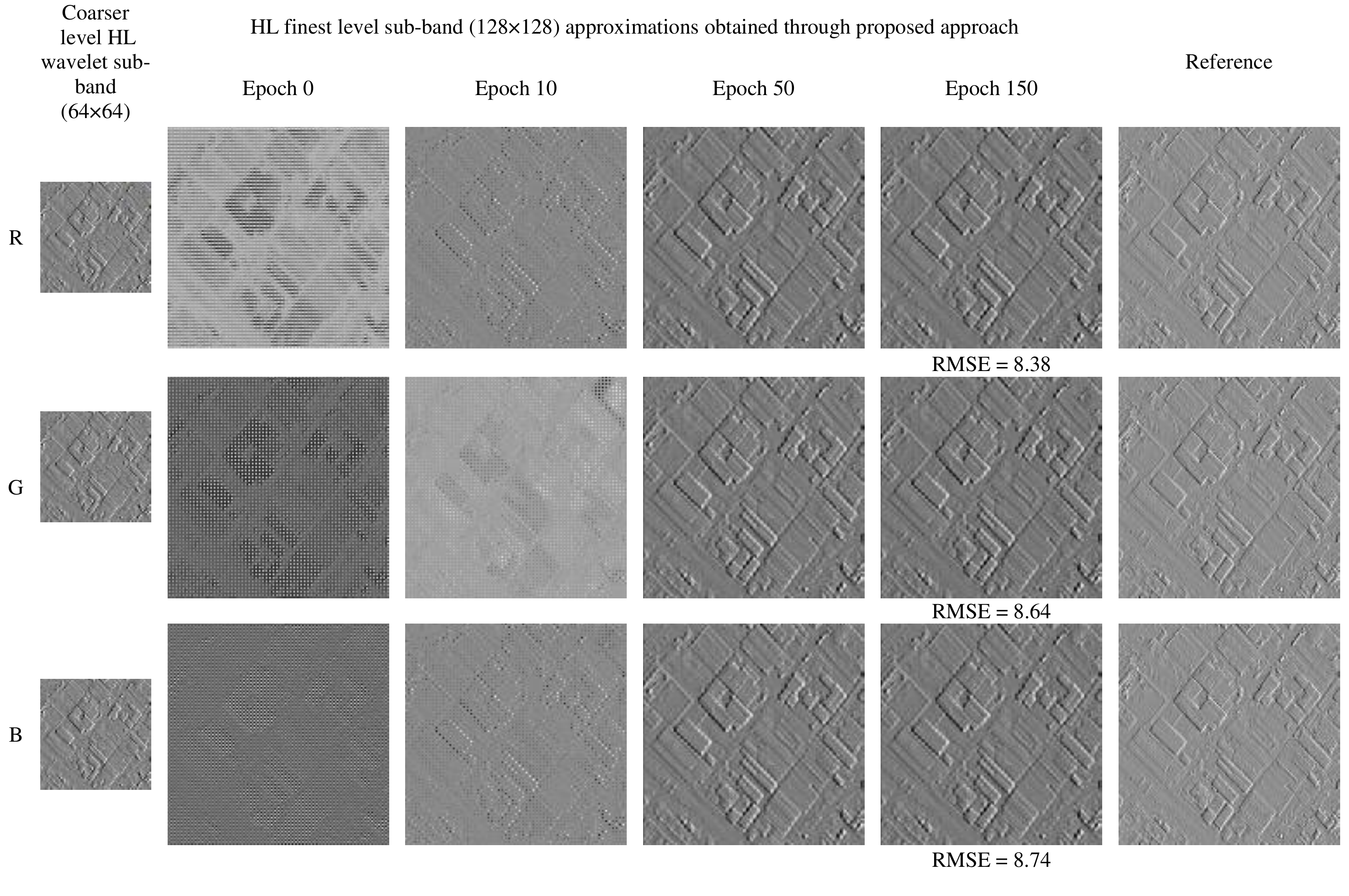}
	\caption{Qualitative results of sub-band approximations associated to HL wavelet sub-band of an image belonging to building category (NWPU-RESISC45 archive).}
\end{figure*}
\begin{figure*}
\centering
		\includegraphics[scale= 0.50]{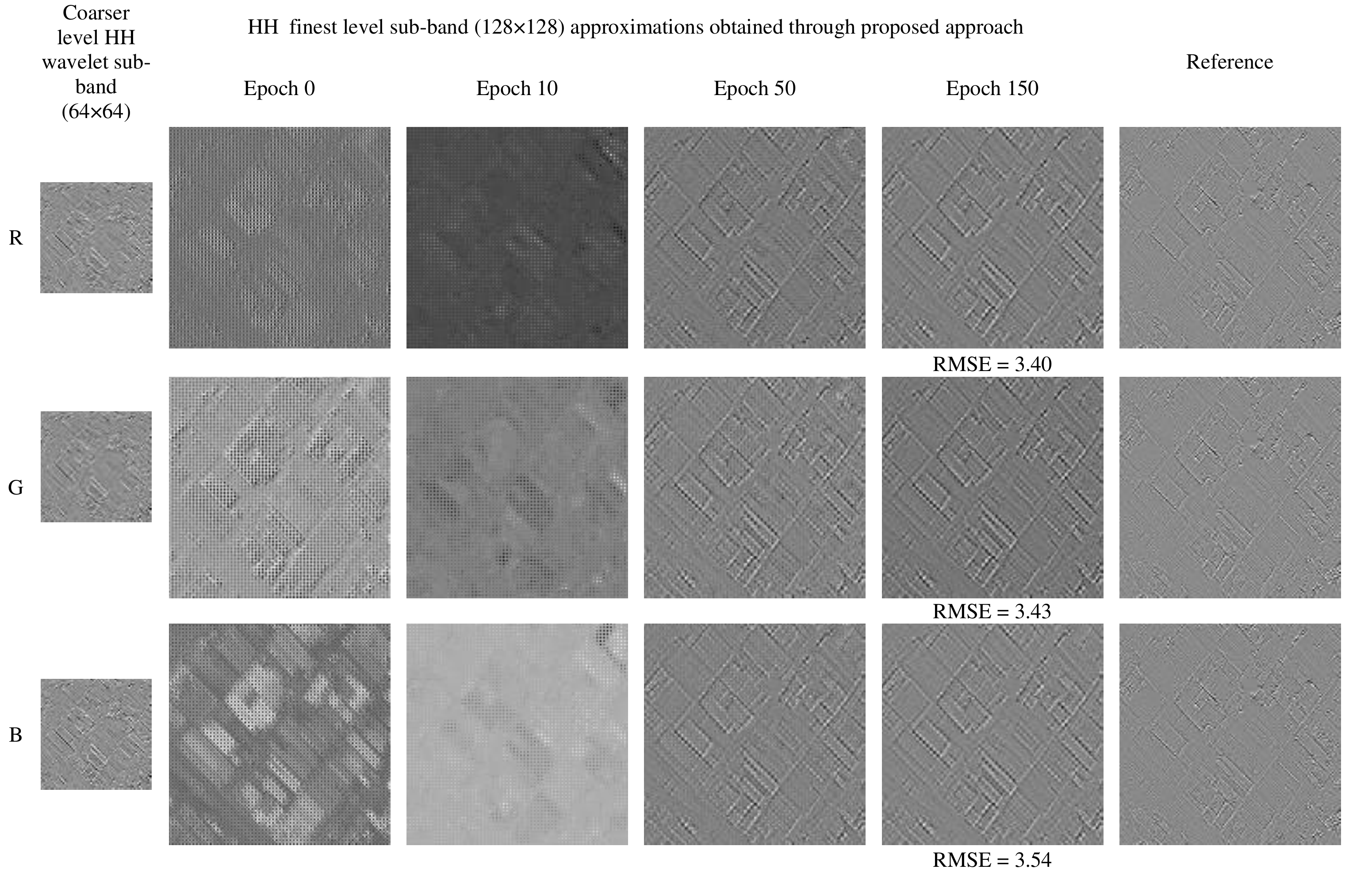}
	\caption{Qualitative results of sub-band approximations associated to HH wavelet sub-band of an image belonging to building category (NWPU-RESISC45 archive).}
\end{figure*}

\section{EXPERIMENTAL RESULTS}
To evaluate the effectiveness of the proposed approach, we performed several experiments to: i) assess the quality of the proposed approximated images compared to the decoded wavelet sub-bands (image); ii) analyze the performance of the proposed approach for scenarios 1 and 2 (mentioned in Section III-A-1 and III-A-2); and iii) compare the performance and computational gain with respect to a standard CNN. To this end, we considered two different cases under scenario 1 and 2 where:
\begin{enumerate}
\item Scenario 1 - the coarsest level wavelet sub-bands are used to approximate the image level information; 
\item Scenario 1 - the coarsest level wavelet sub-bands are used to approximate the intermediate finer level wavelet sub-bands;
\item Scenario 2 - decoded finer level wavelet sub-bands are used to approximate the image level information;
\item Scenario 2 - decoded finer level wavelet sub-bands are used to approximate intermediate finest level wavelet sub-bands.
\end{enumerate}
In the first set of experiments, we assess the qualitative as well as quantitative performances of the proposed approach for scene classification on both NWPU-RESISC45 and AID image archives. 
\subsection{Qualitative Analysis of the Approximated Images}
This subsection provides a qualitative analysis of the images obtained by the proposed approximation approach for the NWPU-RESISC45 archive. Fig. 6-9 show the approximated images obtained for LL, LH, HL and HH wavelet sub-bands for the images of NWPU-RESISC45 archive that are associated to the building class. To qualitatively analyze the efficiency of the proposed approach, we provide the RMSE value between the approximated image and the decoded image. Given a coarser level wavelet sub-band $(64\times64)$ of the image from the NWPU-RESISC45 archive (Scenario 2), one can notice that the proposed approach is efficient to model the finer level wavelet sub-band $(128\times128)$. It converges fast (around Epoch 50) for all the wavelet sub-bands. 
The RMSE values obtained for LL sub-bands are 157.86, 165.62, 166.96 for Red (R), Green (G) and Blue (B) bands, respectively. Transposed convolution used to approximate the finer level wavelet sub-bands (image) introduces a loss to the fine-scale detailed information. This is visible from the LL sub-band finest approximated images $(128\times128)$ (see Fig. 6). The RMSE values for the HL (vertical) sub-bands are 8.38, 8.64 and 8.74 for RGB bands, respectively. In addition, we also notice decreased RMSE values for the detail wavelet sub-bands (which are LH, HL and HH) when compared to the approximation wavelet sub-band (which is LL). Thus, we can see that the transposed convolution efficiently approximates the detail wavelet sub-bands. The same behavior is replicated when AID image archive is used, however, they are not repeated due to space constraints.

\subsection{Quanitative Results of the Proposed Approximation Approach}
\begin{table*}
	\caption{Classification accuracy and computational time for the proposed Approximation approach (NWPU-RESISC45 archive).}
    \centering
	\begin{tabular}{ |c c| c| c| c| c|}
	
	 \hline

     \multicolumn{2}{|c|}{\multirow{2}{*}{\centering Proposed Approximation Approach}} & \multirow{2}{*}{Accuracy (\%)} & \multicolumn{3}{c|}{Computational Time (sec)} \\
     \cline{4-6}

    \multicolumn{2}{|c|}{}  &  & Train & Validation  & Test  \\
     \hline
     
     \multirow{3}{*}{Scenario 1} & \multicolumn{1}{|c|}{\makecell{Approximating image\\($32\times32$) $\rightarrow$ ($64\times64$) $\rightarrow$ ($128\times128$) $\rightarrow$ ($256\times256$)}} & 73.27 & 8770.76 & 6.13 & 5.17\\
     
     \cline{2-6}

       &\multicolumn{1}{|c|}{\makecell{Approximating finest level wavelet sub-bands\\ ($32\times32$) $\rightarrow$ ($64\times64$) $\rightarrow$ ($128\times128$)}} & 74.05 & 6739.87 & 5.28 & 5.68 \\
      
    \cline{2-6}
      &\multicolumn{1}{|c|} {\makecell{Approximating finer level wavelet sub-bands\\($32\times32$) $\rightarrow$ ($64\times64$)}} & 65.42 & 568.99 & 0.38 & 0.51 \\
      
    \hline
    
 \multirow{3}{*}{Scenario 2} & \multicolumn{1}{|c|}{\makecell{Approximating image\\ ($64\times64$) $\rightarrow$ ($128\times128$) $\rightarrow$ ($256\times256$)}} & 80.09 & 8630.20 & 106.37 & 106.51\\
     
     \cline{2-6}

       &\multicolumn{1}{|c|}{\makecell{Approximating finest level wavelet sub-bands\\ ($64\times64$) $\rightarrow$ ($128\times128$)}} & 79.92 & 8393.79 & 102.03 & 101.81 \\
      
    \cline{2-6}
      &\multicolumn{1}{|c|} {\makecell{Approximating image\\($128\times128$) $\rightarrow$ ($256\times256$)}} & 78.54 & 8853.99 & 207.24 & 206.81 \\
      
    \hline
\end{tabular}
\end{table*}

\begin{table*}
	\caption{Classification accuracy and computational time for the proposed Approximation approach  (AID archive).}
    \centering
	\begin{tabular}{ |c c| c| c| c| c|}
	
	 \hline

     \multicolumn{2}{|c|}{\multirow{2}{*}{\centering Proposed Approximation Approach}} & \multirow{2}{*}{Accuracy (\%)} & \multicolumn{3}{c|}{Computational Time (sec)} \\
     \cline{4-6}

    \multicolumn{2}{|c|}{}  & & Train & Validation  & Test  \\
     \hline
     
     \multirow{3}{*}{Scenario 1} & \multicolumn{1}{|c|}{\makecell{Approximating image\\($75\times75$) $\rightarrow$ ($150\times150$) $\rightarrow$ ($300\times300$) $\rightarrow$ ($600\times600$)}} & 74.64 & 14539.87 & 14.26 &14.03 \\
     
     \cline{2-6}

       &\multicolumn{1}{|c|}{\makecell{Approximating finest level wavelet sub-bands\\ ($75\times75$) $\rightarrow$ ($150\times150$) $\rightarrow$ ($300\times300$)}} & 76.92 & 13598.14 & 13.87 &  14.91\\
      
    \cline{2-6}
      &\multicolumn{1}{|c|} {\makecell{Approximating finer level wavelet sub-bands\\($75\times75$) $\rightarrow$ ($150\times150$)}} & 77.34 & 10115.91 & 8.62 & 9.90 \\
      
    \hline
    
 \multirow{3}{*}{Scenario 2} & \multicolumn{1}{|c|}{\makecell{Approximating image\\ ($150\times150$) $\rightarrow$ ($300\times300$) $\rightarrow$ ($600\times600$)}} & 79.91 &14183.46  &253.98 & 279.28\\
     
     \cline{2-6}

       &\multicolumn{1}{|c|}{\makecell{Approximating finest level wavelet sub-bands\\ ($150\times150$) $\rightarrow$ ($300\times300$) }} &79.24 & 13847.33 & 224.36 & 227.34 \\
      
    \cline{2-6}
      &\multicolumn{1}{|c|} {\makecell{Approximating image\\($300\times300$) $\rightarrow$ ($600\times600$)}} & 78.52 & 14964.64  & 326.34 & 331.65 \\
      
    \hline
\end{tabular}
\end{table*}
This subsection presents the classification accuracies and the computational time required by the proposed approach. For the following experiments, the feature extraction and classification steps of the proposed approach have been based on the AlexNet model. Table II reports the performance of the proposed approach (for scenarios 1 and 2) for the NWPU-RESISC45 benchmark archive. Note that the computational time includes the decoding time required for the considered images. From the numbers in Table II associated to Scenario 1, one can notice that the proposed approach employs the coarsest level wavelet sub-bands ($32\times32$) to approximate: i) the image ($256\times256$)  after applying three transposed convolutional layers; ii) the finest level wavelet sub-band ($128\times128$) after applying two transposed convolutional layers; and iii) the finer level wavelet sub-band ($64\times64$) after applying one transposed convolutional layer. As one can observe, approximating the finest level wavelet sub-band  ($128\times128$) achieves the best classification performance when compared to the other two cases. This is due to the fact that when the coarsest level wavelet sub-bands are used to approximate image (which requires three transposed convolution layers), the details of the approximated fine-scaled objects are reduced. Nonetheless, when the finest level wavelet sub-bands ($128\times128$) are approximated (which requires only two transposed convolution layers), we gain in terms of both performance and computational time. Thus, we can conclude that, as the number of layers used for approximation decreases, the performance increases. In the third case, where the finer level sub-bands ($64\times64$) is approximated (using one transposed convolutional layers), the size of the approximated wavelet sub-bands does not provide enough image information for accurate classification. Thus, the resulting classification accuracy is the lowest (i.e. 65.42\%) when compared to the other two cases.

From the numbers in Table II associated to scenario 2, one can see that the proposed approach has used the decoded finer level wavelet sub-bands ($64\times64$) to approximate: i) the image ($256\times256$) after applying two transposed convolutional layers; and ii) the finest level wavelet sub-bands ($128\times128$). In the third case, the proposed approach uses the decoded finest level wavelet sub-bands ($128\times128$) to approximate the image ($256\times256$). As one can see, all the three cases report almost similar classification accuracies with very small differences. However, if we compare the computational times, we can observe that the training time required for approximating the finest level wavelet sub-bands is lower than the time required to approximate the images. The training time required when the finest level wavelet sub-bands are used is 6739.87 sec. In the classification phase, the proposed approach takes 206.81 sec when the image is approximated after decoding two wavelet decomposition levels. In the first case, where the image is approximated using the the finer level wavelet sub-bands ($64\times64$), the required computational time is only 106.51 sec. The overall gain is achieved when the finest level wavelet sub-bands ($128\times128$) is approximated. 

When we compare scenarios 1 and 2 (Table II), we can notice that the proposed approach attains good classification accuracies when the finest level wavelet sub-bands are used. If we analyze the performance of the proposed approach when the finest level wavelet sub-bands ($128\times128$) are obtained, it achieves an accuracy of 74.04\% when the coarsest level wavelet sub-bands are used with a required classification time (i.e. test time) as 5.68 sec. In the other case, approximating the finest level wavelet sub-bands ($128\times128$) after decoding results in 79.92\% classification accuracy with a higher computational time of 101.81 sec. We can observe that the proposed approach obtains accuracy of 74.05\% when only the coarsest level wavelet sub-bands are used with a significantly reduced computational time 5.68 sec. If we perform one level wavelet decoding to obtain the finer level ($64\times64$) wavelet sub-bands, which is used to approximate the finest level wavelet sub-bands ($128\times128$), we notice an increase of 5.87\% in classification accuracy. This shows that the proposed approach achieves reasonable classification performance with the coarsest wavelet sub-bands. Thus, the experimental results demonstrate that the finest level wavelet sub-bands (partially decoded domain) provide sufficient information for an efficient scene classification with reduced computational time.

Table III reports the performance of the proposed approach on the AID benchmark archive. From Table III (Scenario 1 and 2), the proposed approach employs the coarsest and the decoded finer level wavelet sub-bands ($75\times75$) to approximate the finer level wavelet sub-bands (the image itself). While analyzing the part of Table III associated to scenario 1, we can notice that the proposed approach employs the coarsest level wavelet sub-band to approximate: i) the image level information ($600\times600$); ii) the finest level wavelet sub-bands ($300\times300$); and iii) the finer level wavelet sub-band ($150\times150$). From the results, we can observe an accuracy of 77.34\% when we approximate the finest level wavelet sub-band ($150\times150$). This is because the size of the coarsest level wavelet sub-band employed provides sufficient information to approximate finer level sub-bands ($150\times150$) without large information loss with only one transposed convolutional layer. However, when we use two or more transposed convolutional layers to approximate finer level wavelet sub-bands (the image itself), the accuracy is reduced. This is because the coarsest level wavelet sub-bands ($75\times75$) introduce checkerboard artifacts when two or more transposed transposed convolutional layers are included. In addition, the training and classification times required are 10115.91 sec and 9.90 sec, respectively. Thus, it requires minimum decoding, which reduced the additional overhead required before classification.

In the part of Table III associated to scenario 2, the proposed approach employs: i) the finer level wavelet sub-bands ($150\times150$) to approximate the image level information ($600\times600$); ii) the finer level wavelet sub-bands ($150\times150$) finest level wavelet sub-bands ($300\times300$); and iii) the finest level wavelet sub-bands ($300\times300$) to approximate the image ($600\times600$). If we compare the classification accuracies, we can observe that the highest classification accuracy of 79.91\% is achieved when the finer level wavelet sub-bands ($150\times150$) are used to approximate the image ($600\times600$). However, the training time required to approximate the finest level wavelet ($300\times300$) from the finer level wavelet sub-bands is lower when compared to the other two cases and the classification accuracy is 79.24\% which is very close to the highest one. In addition, this last case has also the lowest classification time. From the experimental results, we can conclude that, if the size of the coarsest level wavelet sub-bands is large enough (e.g. as in the case of ($75\times75$) AID archive), the proposed approach requires only one approximation level to achieve an acceptable classification accuracy. 

\subsection{Comparison of the Proposed Approach with a Standard CNN.}
\begin{table*}
	\caption{Classification accuracy and computational time for the proposed approximation approach and a standard CNN (NWPU-RESISC45 archive).}
    \centering
	\begin{tabular}{ |c| c c| c| c| c| c|}
	
	 \hline
    \centering
   
     \multirow{2}{*}{Model}&\multicolumn{2}{c|}{\multirow{2}{*}{Method}} & \multirow{2}{*}{Accuracy (\%)} & \multicolumn{3}{c|}{Computational Time (sec)} \\
     \cline{5-7}
    \centering
    
    &\multicolumn{2}{c|}{}  &  & Train & Validation  & Test  \\
     \hline
     
      \centering  {\multirow{7}{*}{AlexNet}}&\multicolumn{1}{c}{\multirow{3}{*}{\makecell{Proposed Approximation\\ Approach}}} &\multicolumn{1}{|c|}{\makecell{Approximating finest level wavelet sub-bands\\ ($32\times32$) $\rightarrow$ ($64\times64$) $\rightarrow$ ($128\times128$)}} & 74.05 & 6739.87 & 5.28 & 5.68 \\
     
     \cline{3-7}

      \centering {\multirow{7}{*}{}}&\multicolumn{1}{c}{} &\multicolumn{1}{|c|}{\makecell{Approximating finest level wavelet sub-bands\\ ($64\times64$) $\rightarrow$ ($128\times128$)}} & 79.92 & 8393.79 & 102.03 & 101.81 \\

    \cline{2-7}
    
 \centering {\multirow{7}{*}{}}&\multicolumn{1}{c}{\multirow{3}{*}{Standard CNN}} & \multicolumn{1}{|c|}{\makecell{Fully decompressed image \\($256\times256$)}} &80.11  & 7478.89 & 305.13&306.24 \\
     
     \cline{3-7}

      \centering {\multirow{7}{*}{}}&\multicolumn{1}{c}{} &\multicolumn{1}{|c|} {\makecell{Without any decompression\\($32\times32$)}} &  54.01&  314.20& 0.13 & 0.12 \\
        \hline
     
     \centering {\multirow{7}{*}{ResNet50}}&\multicolumn{1}{c}{\multirow{3}{*}{\makecell{Proposed Approximation\\ Approach}}} & \multicolumn{1}{|c|}{\makecell{Approximating finer level wavelet sub-bands\\($75\times75$) $\rightarrow$ ($150\times150$)}} &85.91&\makecell{16953.31} & 15.61 & 16.01 \\

    \cline{3-7}
      \centering \multirow{7}{*}{}& \multicolumn{1}{c}{} &\multicolumn{1}{|c|} {\makecell{Approximating finest level wavelet sub-bands\\($150\times150$) $\rightarrow$ ($300\times300$)}} & 93.98  & 18992.71& 124.32 &125.64\\
      
      \cline{2-7}
      \centering \multirow{7}{*}{}&\multicolumn{1}{c}{\multirow{3}{*}{Standard CNN}} & \multicolumn{1}{|c|}{\makecell{Fully decompressed image \\($600\times600$)}} & 94.85& 17234.51 & 326.63 & 325.98 \\
     
     \cline{3-7}

      \centering \multirow{7}{*}{}&\multicolumn{1}{c}{} &\multicolumn{1}{|c|} {\makecell{Without any decompression\\($75\times75$)}}&76.31  & 763.24 &3.64   & 3.98 \\
      
    \hline

\end{tabular}
\end{table*}
\begin{table*}
	\caption{Classification accuracy and computational time for the proposed approximation approach and a standard CNN (AID archive).}
    \centering
	\begin{tabular}{ |c|c c| c| c| c| c|}
	
	 \hline
    \centering
    
     \multirow{2}{*}{Model}&\multicolumn{2}{c|}{\multirow{2}{*}{Method}} & \multirow{2}{*}{Accuracy (\%)} & \multicolumn{3}{c|}{Computational Time (sec)} \\
     \cline{5-7}
    \centering
    
    &\multicolumn{2}{c|}{}  &  & Train & Validation  & Test  \\
     \hline
     
     \centering {\multirow{7}{*}{AlexNet}}&\multicolumn{1}{c}{\multirow{3}{*}{\makecell{Proposed Approximation\\ Approach}}} & \multicolumn{1}{|c|}{\makecell{Approximating finer level wavelet sub-bands\\($75\times75$) $\rightarrow$ ($150\times150$)}} &77.34&\makecell{10115.91} & 8.62 & 9.90 \\

    \cline{3-7}
      \centering {\multirow{7}{*}{}}&\multicolumn{1}{c}{} &\multicolumn{1}{|c|} {\makecell{Approximating finest level wavelet sub-bands\\($150\times150$) $\rightarrow$ ($300\times300$)}} & 79.24  & 13847.33& 224.36 &227.34\\
      
    \cline{2-7}
    
 \centering {\multirow{7}{*}{}}&\multicolumn{1}{c}{\multirow{3}{*}{Standard CNN}} & \multicolumn{1}{|c|}{\makecell{Fully decompressed image \\($600\times600$)}} & 79.54 & 12582.21 & 412.37 & 422.84 \\
     
     \cline{3-7}

      \centering {\multirow{7}{*}{}}&\multicolumn{1}{c}{} &\multicolumn{1}{|c|} {\makecell{Without any decompression\\($75\times75$)}}&61.91  & 946.23 &7.90    & 8.25 \\

     \hline
     
     \centering {\multirow{7}{*}{ResNet50}}&\multicolumn{1}{c}{\multirow{3}{*}{\makecell{Proposed Approximation\\ Approach}}} & \multicolumn{1}{|c|}{\makecell{Approximating finer level wavelet sub-bands\\($75\times75$) $\rightarrow$ ($150\times150$)}} &84.92&\makecell{17256.34} & 14.32 & 14.13 \\

    \cline{3-7}
      \centering {\multirow{7}{*}{}}&\multicolumn{1}{c}{} &\multicolumn{1}{|c|} {\makecell{Approximating finest level wavelet sub-bands\\($150\times150$) $\rightarrow$ ($300\times300$)}} & 92.24  & b24356.75& 298.26 &299.50\\
      
      \cline{2-7}
      \centering {\multirow{7}{*}{}}&\multicolumn{1}{c}{\multirow{3}{*}{Standard CNN}} & \multicolumn{1}{|c|}{\makecell{Fully decompressed image \\($600\times600$)}} & 93.01 & 21731.25 & 443.91 & 443.12 \\
     
     \cline{3-7}

      \centering {\multirow{7}{*}{}}&\multicolumn{1}{c}{} &\multicolumn{1}{|c|} {\makecell{Without any decompression\\($75\times75$)}}&69.78  & 1231.24 &13.56    & 13.14 \\
      
    \hline
\end{tabular}
\end{table*}
In this subsection, we compare the effectiveness of the proposed approach with: i) a standard-CNN model where full decompression of images is required; and ii) a standard-CNN model that takes as input the coarsest level wavelet sub-bands (which can be obtained from the codestreams of the compressed image). For the following experiments, the feature extraction and classification parts are based on the ResNet50 model. Tables IV and V report the classification accuracies and computational times for the NWPU-RESISC45 and AID image archives, respectively. It is worth noting that during classification the proposed approach requires only the codestreams associated with the coarsest level wavelet sub-bands, whereas the standard-CNN model requires the fully decompressed images. By analyzing the tables one can observe that the computational time required by the proposed approach is significantly reduced when compared to that of the standard-CNN model. In addition, we can also notice that the proposed approach attains almost similar classification accuracies when compared to the standard-CNN model that uses fully decompressed images. On the contrary, if we perform classification using the coarsest level wavelet sub-bands, the classification accuracy is significantly reduced.

By analyzing the AlexNet model results for NWPU-RESISC45 archive (Table IV), we can notice that the classification accuracy obtained by using fully decompressed images with a standard CNN is 80.11\%, with a classification time (i.e. test time) of 306.24 sec. The proposed approach results in a very similar classification accuracy of 79.92\% when only one level of decoding is performed with a lower classification time of 101.81 sec. When the coarsest level wavelet sub-bands ($32\times32$) are used to approximate finest level wavelet sub-bands ($128\times128$), the required classification time is of more than an order of magnitude smaller at the cost of almost 5\% lower classification accuracy. When the coarsest level wavelet-subbands are used in the standard CNN, we obtain the lowest classification accuracy with the lowest classification time. By analyzing the ResNet50 model results for the AID archive (Table IV), the classification accuracy obtained by fully decompressing the images is 94.85\% with a classification time of 325.98 sec. The proposed approach results again in a very similar classification accuracy of 93.98\% by reducing classification time (i.e. test time) to 125.64 sec.

By analyzing the AlexNet model results for NWPU-RESISC45 archive (Table V), we observe that the proposed approach results in a classification accuracy of 77.34\%  when the coarsest level wavelet sub-bands are used, with a classification time of 9.90 sec. When we compare the performance of the proposed approach with the standard-CNN, although the classification accuracy is reduced by 2.20\%, there is a significant gain in terms of the classification time that is reduced to 9.90 sec. Also, it is important to note that the proposed approach reaches a classification accuracy of 79.24\% which is similar to that obtained by the standard-CNN approach that requires fully decompressed images. By analyzing ResNet50 model results for the AID archive (Table V), the classification accuracy obtained by fully decompressing the images is 93.01\%  with a computational time of 444.12 sec. The proposed approach results in a similar classification accuracy of 92.24\% with a computational time 299.50 sec. By analyzing the results, one can conclude that the proposed approach minimizes the computational time considerably when compared to the standard-CNN model. In addition, by using a powerful CNN model like ResNet50, the performance is also improved. However, this is achieved at the cost of increasing the computational time.

\section{DISCUSSION AND CONCLUSION}
In this paper, a novel approach has been presented to perform RS image scene classification in the JPEG 2000 compressed domain by using DNNs. The proposed approach minimizes the amount of image decoding by a DNN, which approximates the finer level wavelet sub-bands from the codestreams associated to the coarser level wavelet sub-bands. To this end, the proposed approach initially takes the codestreams associated to the coarsest level wavelet sub-bands in order to approximate the finer level wavelet sub-bands with a series of transposed convolutional layers. 
The aim of the transposed convolutional layers is to approximate the finer level wavelet sub-bands without requiring to decode the images (in order to obtain the features for scene classification). This significantly reduces the decoding time required for scene classification, which is the dominant aspect while performing scene classification in compressed RS image archives. Then, the features obtained from the finer level wavelet sub-bands are obtained through the convolutional layers. Then, the proposed approach performs scene classification based on the obtained features. During training, in addition to the classification loss, the approximation loss is also optimized that is calculated between the approximated wavelet coefficients and the original wavelet coefficients. 
Thanks to the approximation of finer-level wavelet sub-bands, the time required to decode the images is considerably minimized. 

Experimental results in terms of scene classification accuracy and computational gain on two benchmark archives demonstrate the effectiveness of the proposed approach. This is mainly related to the significant reduction of the decoding time associated with the use of a large amount of compressed images. Since there is a trade-off between the computational gain and the classification accuracy based on the number of transposed convolutional layers, one can always choose the number of layers depending on the requirements in computational time and accuracy. The qualitative images obtained from the approximations show that the proposed approach efficiently operates only with the original coarsest level wavelet coefficients as input source. The results obtained from the experiments demonstrate the ability of the proposed approach:
\begin{enumerate}
\item To accurately perform image scene classification in the JPEG 2000 compressed domain.
\item To significantly improve the computational gain by minimizing the required amount of decompression compared to the existing scene classification methods (which operates on uncompressed images).
\end{enumerate}

In view of the growth of RS big data archives, this work introduces a research direction for operating scene classification with DNNs directly on the compressed archives. Note that the proposed approach is not limited to JPEG 2000 compressed archives but can be directly applied to any image archive that considers wavelet based compression approach. In addition, the introduced concept can be adapted to be used in the framework of other compression algorithms by properly modifying the technique used for approximating the compressed domain features. As a future development of this work, we plan to explore scene classification in the context of GANs in the compressed domain. Moreover, we plan to study the development of specific models that can extract features within a deeply compressed domain.
\begin{table*}[!t]
	\caption{List of Symbols used in this paper }
    \centering
	\begin{tabular}{ |c| c| }
	
    \hline
    \centering
     Symbol  &   Description   \\
     
     \hline
     \centering $\textbf{X}$ & an archive that consists of \textit{N} JPEG 2000 compressed images  \\
     
     \hline
      \centering $X_i$  & $i^{th}$ image in the archive $\textbf{X}$ \\
      
     \hline
     \centering $\textbf{Y}$  & set of class labels associated with archive $\textbf{X}$ \\
     
     \hline
      \centering $y_i$  & $i^{th}$ class label in set $\textbf{Y}$ \\
      
     \hline
     \centering $L$  & number of wavelet decomposition levels used in archive $\textbf{X}$ \\
     
     \hline
     \centering $G^L$ & the coarser approximation, horizontal, vertical and diagonal sub-bands of an image $X_i$ at $L^{th}$ wavelet decomposition level\\
    
     \hline
     \centering $A^{L-1}$ & the finer approximation, horizontal, vertical and diagonal sub-bands of an image $X_i$ at ${L-1}^{th}$ wavelet decomposition level\\
    
     \hline
     \centering \textbf{C} & sparse matrix\\
    
     \hline
     \centering S & stride used in the DNN\\
    
     \hline
     \centering P & padding used in the DNN\\
    
     \hline
     \centering $A^{L-1}_{size}$ & approximated wavelet sub-band size\\
    
     \hline
     \centering $A^L_{size}$ & coarser wavelet sub-band size\\
    
     \hline
     \centering $k$ & considered kernel\\
    
     \hline
     \centering $\mathcal{L}_{total}$ & total loss\\
     
     \hline
     \centering $M$ & number of rows in the considered wavelet sub-band\\

     \hline
     \centering $N$ & number of columns in the considered wavelet sub-band\\
    
     \hline
     \centering $\mathcal{L}_{approximation}$ & approximation loss\\
    
     \hline
     \centering $\mathcal{L}_{classification}$ & classification loss\\
    
     \hline
     \centering $\hat{y}_{i}$ & predicted class label\\
    
     \hline
     \centering $D^i$ & decoded wavelet sub-band\\
    
     \hline
\end{tabular}
\end{table*}

\appendix
A list of notation and symbols used throughout this paper is provided in Table VI.

\bibliographystyle{IEEEtran}
\bibliography{ref.bib}

\end{document}